\documentstyle[preprint,aps]{revtex}
\draft
\begin{document}
\title{Current Distribution in the Three-Dimensional\\
Random Resistor Network at the Percolation Threshold}
\author{G.\ George Batrouni}
\address{HLRZ Forschungszentrum, D--52425 J{\"u}lich, Germany}
\author{Alex Hansen}
\address{Institutt for fysikk, Norges tekniske h{\o}gskole,
N--7034 Trondheim, Norway}
\author{Brond Larson\footnote{Current address: Silicon Graphics Computer
Systems, 1 Cabot Road, Hudson, MA 01749, USA.}}
\address{Thinking Machines Corporation, 245 First Street, Cambridge
MA 02142, USA}
\date{\today}
\maketitle
%%%%%%%%%%%%%%%%%%%%%%%%%%%%%%%%%%%%%%%%%%%%%%%%%%%%%%%%%%%%%%%%%%%%
\begin{abstract}
We study the multifractal properties of the current distribution of the
three-dimensional random resistor network at the percolation threshold.
For lattices ranging in size from $8^3$ to
$80^3$ we measure the second, fourth and sixth moments of the
current distribution, finding {\it e.g.\/} that $t/\nu=2.282(5)$ where $t$
is the conductivity exponent and $\nu$ is the correlation length exponent.
\end{abstract}
%%%%%%%%%%%%%%%%%%%%%%%%%%%%%%%%%%%%%%%%%%%%%%%%%%%%%%%%%%%%%%%%%%%
\section{Introduction}
\label{sec1}
Quite surprisingly, it was found in the mid-eighties that dynamic phenomena
on fractal structures often were controlled not by one or two relevant length
scales, but rather an infinite hierarchy of such length scales
\cite{hjkps86}.
One of the prime examples of this phenomenon is the current
distribution in the random resistor network at the percolation threshold
\cite{rtbt85,arc85,rtt85,arc86}.  In spite
of the large effort which was invested to understand how such an
infinite hierarchy could appear, {\it e.g.\/} through studying hierarchical
structures yielding to analytic calculations \cite{arc85,arc86},
no satisfactory general
explanation was found.  The large numerical effort that was invested at that
time on the random resistor network focussed on two dimensions.  The reason
for this was that three-dimensional networks were essentially out of reach
of the computational power available at the time.  The aim of the
present work is to establish that, as in two dimensions, the current
distribution in three dimensions is multifractal and to determine the
corresponding scaling exponents with high presicion.  In addition to their
theoretical interest, these results are of importance in making contact with
experimental studies on three-dimensional conductor-insulator mixtures
\cite{lsncg86,snlg86,rm87}, and microemulsions
\cite{l79,ls80,bskh85,eht86,c88}.
To achieve this goal, we made use of the latest developments in iterative
solvers and massively parallel computers.

In section \ref{sec2} we present the model and the method of solution.
In section \ref{sec3} we discuss the current distribution through the behaviour
of the
moments and their exponents, in addition to examining the statistical
fluctuations of the  moments.
%---------------------------------------------------------
\section{Model and Numerical Solution}
\label{sec2}
We consider a three-dimensional cubic lattice of size $L^3$ with
periodic boundary conditions in the $x$ and $y$ directions. For the
$z$ direction the boundary conditions are as follows: At $z=1$ and
$z=L$ we place two plates with a constant potential difference set at
a value of 1. Therefore, the length of the lattice in the $z$
direction is $(L-1)$, while in the $x$ and $y$ directions the length
is $L$. This geometry was chosen because the data layout becomes
optimal on the Connection Machine CM5 which we used for our
computations.

All bonds are visited and, with a probability $p$, a resistor is
placed. All resistors have the same resistance, 1. We set $p$ equal to
the bond percolation threshold for the cubic lattice, $p_c=0.2488$, as
determined by Stauffer {\it et al.\/} \cite{saa94}. After all the
bonds have been visited, a (parallel) cluster finding algorithm is
applied to determine if there is a spanning cluster that connects the
two plates. If no such cluster exists, the procedure is repeated until
one is obtained.  This cluster finding algorithm is very fast but only
determines if there is such a cluster, not its exact geometry.

The equations to be solved are the usual current equations (Kirchhoff's
equations) on the lattice which can be easily solved using the
conjugate gradient or related iterative algorithms \cite{bh88}. We wrote the
program in CM Fortran and used the iterative solvers in the
Connection Machine Scientific Software Library (CMSSL) which contains
thirteen of them. Which one to use, depends on the particular matrices
one is dealing with. For this problem we found that the
quasi-minimized CGS (QCGS) \cite{t92} algorithm performs very well.

Our stopping condition is for the residual to be less than $10^{-12}$,
which for the biggest lattices gave a true accuracy of about
$10^{-9}$. We estimated this by calculating the conductivity of each
realization in two ways: 1) Calculate the total current crossing an
$XY$ plane at $z=L/2$, which, knowing the potential difference (=1),
gives the conductivity, and 2) calculate the second moment of the
currents. This is again equal the conductivity as the externally applied
potential difference is 1. On the biggest lattices, for which the accuracy
at which we can determine the currents is the lowest,
they agreed to within $O(10^{-9})$. As it is too time consuming compared to
solving the Kirchhoff equations, we made no effort to identify the bonds
that did not belong to the spanning cluster. We simply solved for the
currents keeping the disconnected bonds and dangling ends in the system.
This made it impossible to get an accurate count of the number of
current-carrying bonds. We will, therefore, give results for the
second, fourth, and sixth momemts of the currents, leaving out the zero'th
moment. The number of realizations for $L=8$, 16, 32, 48, 64, and 80 were
12000, 9198, 2765, 1120, 1913, and 953 respectively.
%---------------------------------------------------------
\section{Current Distribution}
\label{sec3}
Solving for the currents in the bonds allows us to calculate the $n$th
moment of the current distribution, given by
\begin{equation}
\label{eq1}
\langle i^n \rangle_V = {1 \over N_R} \sum_R \sum_{k(R)} i^n_{k(r)},
\end{equation}
where $R$ denotes the realization, $N_R$ is the number of these
realizations, and $i_{k(R)}$ is the current in the $k$th bond in
realization $R$. These moments are calculated in the constant voltage
ensemble where the potential difference is kept constant
realization to realization. We expect the moments of
the currents to scale as
\begin{equation}
\label{eq2}
\langle i^n \rangle_V \sim L_z^{x(n)},
\end{equation}
where $L_z=L-1$ is the length of the system in the $z$ direction, and $x(n)$
is the exponent of the $n$th moment. Figures \ref{fig1}a, \ref{fig2} and
\ref{fig3} show, on a
log-log scale, the second, fourth and sixth moments respectively. The
second moment measures the total dissipated power, and since the
externally applied voltage difference is 1, it directly gives the
conductance of the network. So, Figure \ref{fig1}a gives the scaling of the
conductance as a function of $L_z$, giving the exponent
$x(2)=1.282(5)$. Figure \ref{fig1}b shows the scaling of the {\it variance\/}
of
the conductance distribution as a function of $L_z$. We see that it
scales with the same exponent as the conductance itself. The same is
true for the other moments. Thus, the relative fluctuations neither
grow nor decrease with lattice size.  The scaling
of the {\it conductivity\/} with $L_z$ is obtained by simply dividing
the results for the second moment by $L_z$ (since this is the conductance)
giving $t/\nu=2.282(5)$, where $t$ is the conductivity exponent and $\nu$
is the corrrelation function exponent. Our value for $t/\nu$ is in agreement
with, but more precise than, the value determined by Gingold and Lobb
\cite{gl90} who obtained $t/\nu=2.276(12)$.  With $\nu=0.88$,
we therefore have that $t=2.01$.  The experimental values reported for this
exponent ranges from 1.2 to 2.1 for the measurements based on microemulsions
\cite{l79,ls80,bskh85,eht86,c88}, and for the measurements based on
conductor-insulator mixtures, the values $2.0\pm 0.2$ \cite{lsncg86},
$1.85\pm 0.25$ \cite{snlg86}, and $1.6\pm 0.1$ \cite{rm87}.

The exponent of the fourth moment is related to the scaling of the Nyquist
noise of the random resistor network through the fluctuation-dissipation
theorem as shown by Rammal {\it et al.\/} \cite{rtt85}.
Thus, as for the second
moment, the fourth moment is related to a macroscopic quantity and
therefore is of direct experimental interest.  However, no relation between
the sixth moment of the current distribution and a directly measurable
quantity has been identified.  We determine $x(4)=3.920(6)=$ and
$x(6)=6.477(10)$.

If $y(n)$ is the exponent of the $n$th moment in the constant current
ensemble, we have $\langle i^n \rangle_c \sim L_z^{y(n)}$. It is then
easy to show that $y(n)=x(n) - nx(2)$.  This way, we can easily change
between the two ensembles. Figure \ref{fig4}
shows $y(n)$ as a function of $n$,
where we see that it behaves in the classic multifractal
way \cite{arc86,h90}: The three exponents we show do not fall on a
straight line and as $n$ increases they approach a constant that must
equal $1/\nu$ \cite{rtbt85,bhn87}.
This is shown by the dashed line in the figure. To show
this relation even more clearly, we plot in Figure \ref{fig5} $y(n)$
{\it versus\/}
$1/n$, and where we use for $y(\infty)$ the best value we found for
$\nu=0.88$, \cite{amah90} and references therein. The dashed
line is merely to guide the eye. It
connects the point at $1/n=0$ to that at $1/n=0.5$. We see that the
exponents for the fourth and sixth moments are in agreement with this line
although the sixth moment is starting to lose precision. These $y(n)$ plots
demonstrate the multifractality of the current distribution in this network,
but the convergence to $1/\nu$ as $n\to \infty$ is rather slow, especially
when compared to the two-dimensional case \cite{h90}.

The sample to sample fluctuations of the values of the conductivity
($G$) and the fourth moment (related the Nyquist noise strength, let
us call it $K=\langle i^4\rangle_V$) also yield interesting
information about the system \cite{hma90}.  We have demonstrated in
Figure \ref{fig1}b that the variance of the conductance (from
sample-to-sample fluctuations) scales as the conductance itself with
respect to $L_z$.  Exactly the same behaviour is observed for the
variances of the higher moments.  Therefore, the distributions of
$G/\langle G\rangle$ from different size systems will collapse onto a
single distribution. The same is true for $K/\langle K\rangle$. To
characterize these distributions, we examined on semi-log scale these
distributions against $(G-\langle G\rangle)^2/\langle G\rangle^2$ and
$(K-\langle K\rangle)^2/\langle K\rangle^2$, respectively. Such plots
should yield straight lines for Gaussian distributions. This way we
found the distributions not to be normal, and a similar procedure
showed them not to be lognormal. In Figures \ref{fig6} and \ref{fig7},
we show the distributions of $G/\langle G\rangle$ and $K/\langle
K\rangle$ for different lattice sizes on semi-log scale plotted
against $G/\langle G\rangle$ and $K/\langle K\rangle$.  For the larger
values, we find in both plots straight lines which indicate exponetial
distributions of the form $N(G/\langle G\rangle) \sim \exp(-2G/\langle
G\rangle)$ and $N(K/\langle K\rangle)\sim \exp (-1.3K/\langle
K\rangle)$, respectively.

%---------------------------------------------------------
\section{Conclusions}
\label{sec4}
By using a combination of efficient algorithms and a massively
parallel computer, we were able to do a high precision study of the
distribution and moments of currents in a three dimensional network at
the percolation threshold.

Our result for the conductivity exponent is in agreement with but more
precise than previous values. In addition we evaluated the exponents
for the fourth (related to {\it 1/f\/} and Nyquist noise) and sixth moments.
The values
we have found support the notion of a multifractal current distribution
for the three dimensional network.

We have furthermore studied the sample-to-sample fluctuations and the
distributions of the second (conductivity) and fourth (noise) moment
of the current distribution.  We find that the relative fluctuations
scale as the moments themselves with lattice size, and that the
underlying statistical distributions appear to be exponential rather
than Gaussian. Furthermore, the distribution of conductivities at the
percolation threshold has proven to be an important ingredient in the
formulation of scaling theories for the optical properties (ac
conductivity, reflectivity, transmittivity) of two-dimensional systems
such as semiconductor metal films\cite{yagil} and metal-insulator
composites\cite{noh}. Similar scaling theories in three dimensions,
when constructed, will also need the distribution of the
conductivities for three-dimensional systems at the percolation
threshold, which we have presented in this paper.
%---------------------------------------------------------

\acknowledgements
We thank B.\ Kahng, S.\ Roux, P.\ Tamayo, and B.M.\ Thornton for
helpful discussions. We also thank Institut de Physique du Globe
(Paris), GMD (Bonn), and Thinking Machines Corporation for their
support through generous allocation of time on their CM5 machines.

%%%%%%%%%%%%%%%%%%%%%%%%%%%%%%%%%%%%%%%%%%%%%%%%%%%%%%%%%%%%%%%%%%%

%%%%%%%%%%%%%%%%%%%%%%%%%%%%%%%%%%%%%%%%%%%%%%%%%%%%%%%%%%%%%%%%%%%
\begin{figure}
\caption{\label{fig1} a) The second moment of the current ({\it i.e.\/}
the conductance $G$) scales as a power of the system size $L_z$,
with an exponent equal to $-1.282(5)$. b) The variance of the conductance
distribution scales exactly as the conductance itself with the system size,
$L_z$.}
\end{figure}

\begin{figure}
\caption{\label{fig2} Power law scaling of the fourth moment as a
function of the system size $L_z$. The exponent is $-3.920(6)$.}
\end{figure}

\begin{figure}
\caption{\label{fig3} Power law scaling of the sixth moment as a
function of the system size $L_z$. The exponent is $-6.477(10)$.}
\end{figure}

\begin{figure}
\caption{\label{fig4} $y(n)$, the exponent of the $n$th moment in
the constant current ensemble, as a function of $n$. As $n$ increases,
$y(n)$ approaches the dashed line which is given by $1/\nu$.}
\end{figure}

\begin{figure}
\caption{\label{fig5} Same as Figure 4, but shows $y(n)$ as a function
of $1/n$. This shows very clearly that as $n\to \infty$, $y(n)\to 1/\nu$.
The dashed line is a guide to the eye and simply connects the points at
$1/n=0$ and $1/n=0.5$.}
\end{figure}

\begin{figure}
\caption{\label{fig6}
Histogram of $G/\langle G\rangle$ on semilog scale.
It shows that the distribution for larger values of $G/\langle G\rangle$ is
exponential.}
\end{figure}

\begin{figure}
\caption{\label{fig7}
Histogram of $K/\langle K\rangle$ on semilog scale.
It shows that the distribution for larger values of $K/\langle K\rangle$
is exponential.}
\end{figure}
%%%%%%%%%%%%%%%%%%%%%%%%%%%%%%%%%%%%%%%%%%%%%%%%%%%%%%%%%%%%%%%%%%%
 \end{document}